# Neuro-OSVETA: A Robust Watermarking of 3D Meshes

Bata Vasic[*], Nithin Raveendran[+], Bane Vasic[+], *Fellow IEEE*
[*]Electronic Department, Faculty of Electronic Engineering, University of Nis , Serbia
[+]Electrical and Computer Engineering Department, University of Arizona, Tucson

## Abstract

Best and practical watermarking schemes for copyright protection of 3D meshes are required to be blind and robust to attacks and errors. In this paper, we present the latest developments in 3D blind watermarking with a special emphasis on our Ordered Statistics Vertex Extraction and Tracing Algorithm (OSVETA) algorithm and its improvements. OSVETA is based on a combination of quantization index modulation (QIM) and error correction coding using novel ways for judicial selection of mesh vertices which are stable under mesh simplification, and the technique we propose in this paper offers a systematic method for vertex selection based on neural networks replacing a heuristic approach in the OSVETA. The Neuro-OSVETA enables a more precise mesh geometry estimation and better curvature and topological feature estimation. These enhancements result in a more accurate identification of stable vertices resulting in significant reduction of deletion probability.

## I. INTRODUCTION

We are witnessing a neural network revolution, utilizing deep learning techniques impacting almost every technological niche. Advancements in computation and connectivity has enabled and propelled this transformation. Interestingly, the first revolutionary ideas using neural network actually appeared in the fields of the image classification [1], [2], and object recognition [3]. Image watermark was also recognized as an important area and widely studied in many recent research papers [4], [5], and [6]. Copyright protection of neural networks themselves was also enabled by the watermark techniques [7]. Despite the initial ideas from the last century, scientific interest in using artificial intelligence to three-dimensional (3D) objects and their classification and recognition is still in its infancy. The reason is partially attributed to non-unified representation and description of 3D objects, as well as the complexity of the analysis of geometric networks, but also relatively undeveloped tools and Internet engines for their use. Nevertheless, the past few years have witnessed a progress in this area, and initial results.

As in the field of image processing, the 3D object recognition started with the 3D shape description research. Similarly, to the contribution trends in two-dimensional (2D) case mentioned above, the direction of 3D curve research employs neural networks. More recently, Convolutional Neural Network (CNN) became of active interest in 3D object recognition, shape analysis and 3D scene synthesis [8], [9], and [10]. However, according to the best of our knowledge there are no results in watermarking using techniques from artificial intelligence. The only one attempt [11] - while presents an original idea - lacks mathematical or methodological explanation of the use of neural networks. Moreover, the enthusiastic attempt of authors to describe 3D geometry with only one featured vector did not ensure a inserted watermark robustness.

In this paper, we consider the problem of watermarking 3D meshes using neural networks. As we demonstrate, this approach is natural for this problem. Mathematical formulation of a criterion for selecting vertices in the 3D mesh where the watermark information will be hidden (hosted) is complicated and involves many parameters a priory unknown or data dependent, which is exactly the situation which neural networks handle very well.

Our method, NEURO-OSVETA employs neural network in the vertex extraction algorithm OSVETA [13]. It enables more precise mesh geometry estimation and better curvature and topological feature estimation. These enhancements result in a more accurate identification of stable vertices resulting in significant reduction of the vertex deletion probability in the process of mesh simplification.

The rest of the paper is organized as follows. The problem definition is summarized in Section II. In Section III we explain the relevant OSVETA notations and definitions as well as list of all used geometrical and topological criteria for 3D mesh characterization. Section IV presents theoretical principles of the neural network construction. Also, in this section we talk about our NN learning, back-propagation idea and optimization methods. Section V explains our proven watermarking scheme that ensures additional watermark robustness. Finally, in Section VI we show prelaminar results of vertex deletion probability and error probability performance.

## II. PROBLEM DEFINITION

The first paradigm in 3D watermarking is an ability to obtain robustness of the watermark data to most common geometrical and topological transformations, but also ensuring adequate scheme to define rules for watermark retrieval. To ensure blindness of embedded data, the retrieving algorithm requires some unique topological representation of 3D model and its primitives. However, due to the 3D model representation, we typically deal with an undefined order of mesh primitives and thus with synchronization issues in the watermark data retrieving task. Assuming we provided the above conditions, the second challenge is obtain a reliable scheme to select hosts for embedding watermarking data. Hence, the most important feature of chosen hosts is robustness to transformations, including affine transformations and the very complex optimization and simplification operations.

To generalize all robustness and blindness watermark requirements, we need to ensure definition of the importance of primitives that can be used as hosts for watermark data embedding. Thus, all the selected primitives would be invariant to all transformation and deformation processes even to destructive decimation algorithms. Selected primitives should be capable to hold all watermark information including synchronization requirement in the watermarking retrieval task. Such primitives should be also defined as invariant to topological reordering. Actually, vertices in the 3D mesh that are participating in the shape creation satisfy all above requirements. Thus, only geometrically important set of vertices can be used as watermark host vector and our goal is defining such set of the host vertices. The importance of host vertices is defined as an invariance to its deletion in optimization and simplification process.

## III. ORDERED STATISTICS VERTEX EXTRACTION AND TRACING ALGORITHM

In order to define main neural perception of the shape, we start with a discussion of the 3D shape characterization. However, defining important vertices, which form a simple and computer-recognizable 3D shape, is not a trivial process. Thus, it is necessary to first define the problems of discrete curvature estimation, but also note differences from the known methods.

*A. Algorithm definition and notation*

The OSVETA consists of three steps: i) defining and ranking the vertex assessment criteria, ii) accurate curvature evaluation, and computing its characteristic features, and iii) tracing the importance of extracted vertices in relation to the mesh topology. More precisely, if $M(V,F)$ is a mesh of a given 3D surface, where $V$ and $F$ are respectively vectors of vertices and topological connections between them, the algorithm produces as a result the two vectors: $\mathbf{s}$, the vector of vertex stabilities arranged in a decreasing order, and $\mathbf{q}$, the vector of corresponding indices. The mesh vertices are ordered with respect of decreasing stability form the vector $V_o = \mathbf{v_i}$, and the length-$L$ vector $\mathbf{p}$ is obtained by taking the first $L$ elements from $V_o = \mathbf{v_i}$.

The OSVETA gives good results even for small number of descriptors, but there are more descriptors that can be considered for mesh segmentation [13]. In this moment, to make idea clear and, we would keep attention to considering only descriptors from the given Table 1 [13].

**Table 1** OSVETA ASSESSMENT CRITERIA AND THEIR RATING

| $N^0$ | Criterion | Description | Rate |
|---|---|---|---|
| 1 | $\psi_{min} \geq 0$ | Positive minimal dihedral | 1.0 |
| 2 | $\Theta < 2\pi$ | Small Theta angle | 1.0 |
| 3 | $\kappa_{G1} > 0$ | Positive Gaussian curvature | 1.0 |
| 4 | $\psi_{max} \geq 0$ | Positive maximal dihedral | 0.9 |
| 5 | $\theta > 2\pi$ | Big Theta angle | 0.8 |
| 6 | $\kappa_G < 0$ | Negative Gaussian curvature | 0.8 |
| 7 | $\kappa_{G1} < 0$ | Negative Gaussian curvature | 0.7 |
| 8 | $\kappa_G > 0$ | Positive Gaussian curvature | 0.4 |

This list shows notations of variables, functions and conditions for a given feature values, but also rates of all descriptors that will be later used as an initial weights values in the backpropagation process. In Figure 1, we show the main angles and areas connected to vertex of interest $v_i$.

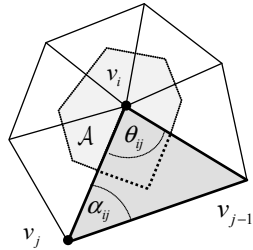

**Figure 1.** Illustration of relevant features and variables at the vertex of interest $\mathbf{v}_i$

*B. Discrete Gaussian curvature descriptors*

The main issue in conditions that generalize computations for 3D meshes is the accuracy for discrete mesh computations. As we can notice from table, *Gaussian curvature* is one of the most relevant shape descriptors. Thus the brief discussion of both methods for discrete Gaussian curvature estimation follows.

*1) Differential geometry method*

We know from [14] that for manifold surface $M$ in $\mathbb{R}^3$, and each point on the given surface, one can locally approximate the surface by its tangent plane that is orthogonal to the *normal vector* **n**. **K** is defined as the normal curvature of the curve that belongs to both the surface itself and the plane containing both **n** and unit direction **e** in the tangent plane. We define the *mean curvature* $\kappa_H = (\kappa_1 + \kappa_2)/2$ as the average value of the two *principal curvatures* $\kappa_1$ and $\kappa_2$ of the surface $M$, The Gaussian curvature $\kappa_G$ is defined as the product of the two principle curvatures, i.e., $\kappa_G = \kappa_1 \kappa_2$. For a given point $\mathbf{v}_i$ of discrete surface $M$ the mean curvature normal (Laplace-Beltrami operator) $\mathrm{K}(\mathbf{v}_i) = 2\kappa_H(\mathbf{v}_i)\mathbf{n}(\mathbf{v}_i)$ gives both the mean curvature $2\kappa_H(\mathbf{v}_i)$ and unit normal $\mathbf{n}(\mathbf{v}_i)$ at the vertex $\mathbf{v}_i$. Gaussian curvature of a discrete surface that depend only on a vertex position and angles of adjacent triangles respectively are given as:

$$\kappa_G(\mathbf{v}_i) = \frac{1}{\mathcal{A}}\left(2\pi - \sum_{j=1}^{\#f}\theta_{ij}\right) \tag{1}$$

A number of adjacent triangular faces at the point $\mathbf{v}_i$ is $\#f$, and $\theta_{ij}$ is the angle of *j*-th adjacent triangle at the point $\mathbf{v}_i$. The area of the first ring of triangles around the point $\mathbf{v}_i$ is $\mathcal{A}$ [15].

*2) Fitting a quadric method*

The idea of Fitting Quadric Curvature Estimation method is that a smooth surface geometry can be locally approximated with a quadratic polynomial surface. The method fits a quadric to point in a local neighborhood of each chosen point in a local coordinate frame. The curvature of the quadric at the chosen point is taken to be the estimation of curvature for the discrete surface. For a simple quadric form $z' = ax'^2 + bx'y' + cy'^2$ procedure is given by: i) Estimation of surface normal **n** at the point **v** by one of two ways: simple or weighted averaging, or finding a least squares fitted plane to the point and its neighbors; ii) Positioning of a local coordinate system $(x', y', z')$ at the point **v** and aligning axis $z'$ along the estimated normal. McIvor and Valkenburg [16] suggest aligning of the $x'$ coordinate axis with a projection of the global $x$ axis onto the tangent plane defined by **n**. If we use the suggested improvements and fit the mapped data to extended quadric: $\hat{z} = a'\hat{x}^2 + b'\hat{x}\hat{y} + c'\hat{y}^2 + d'\hat{x} + e'\hat{y}$, and solve the resulting system of linear equations, we can compute both principal curvatures $\kappa_1$ and $\kappa_2$. Then, we have estimation for Gaussian curvature:

$$\kappa_{G1} = \frac{4a'c' - b'^2}{\left(1 + d'^2 + e'^2\right)^2} \tag{2}$$

## C. Topological features

The second group of topological features that we can use in consideration is a set of features that characterize salience of regions and resistance to transformations. Since the optimization and simplification process destructively affect the object geometry, some angles, vertices, edges and faces are classified as risky primitives (e.g. vertices that are connected by collinear edges, boundary vertices, vertices, edges and faces in flat and smooth areas) and even topological errors (isolated vertices independent in space, vertices which belong to only one boundary edge, complex vertices and edges that join more than two faces, crossed edges without mutual vertex).

However, some of topological features are highly important for shape description (e.g. both, minimal and maximal dihedral angle between two faces at the same edge: $\psi_{max}$ and $\psi_{min}$ respectively.

Our novel technique for 3D shape characterization uses neural network (NN) architectures that are capable to operating on essential geometric and topological representations of 3D meshes. We now describe the NN architecture focusing on neural OSVETA implementation.

## IV. Neural OSVETA

Contrary to 2D image classification CNNs, data sparsity and computation cost of 3D convolution severely constrains the volumetric representation of 3D meshes and points. In the watermarking purpose the result of this classification will be loss of vertex information, and also all the watermark data. Our aim is to define the main characteristics of 3D objects given by their basic representation and appropriate relation of this feature definitions and input representation. Following the principles of Feature-based Neural Networks and including all relevant 3D mesh features, we design a feature learning framework that directly uses vector of all vertices as the NN inputs.

### A. Neural network architecture

An input neuron matrix $I = \{\mathbf{v}, \mathbf{f}\}$ is represented as a set of the following vectors: i) the vector of vertices $\mathbf{v} = \{\mathbf{v}_i \mid i = 1, ...., n\}$, where each vertex is a vector of its $\mathbf{v}(v_x, v_y, v_z)$ Euclidean coordinates, and ii) the topological channel vector $\mathbf{f}(i) = \{f_j \mid j = 1, ..., m\}$ with all vertex connections that represent faces in the given 3D mesh. The diagram in Figure 2. shows the schematic architecture of the proposed neural network.

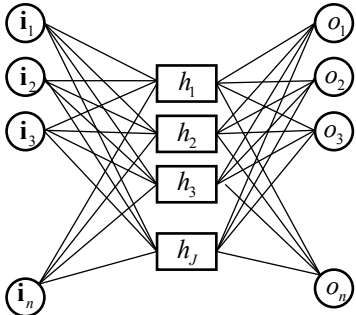

**Figure 2.** The neural network architecture: Input neurons represent elements of the vector **i** that contain geometrical and topological data **v** and **f** respectively. The hidden layer is a set of features that describe shapes inside the 3D mesh (Section II )

The hidden layer neurons represent the geometrical and topological descriptors and conditions given in the Table **1** according to our unique requirements. Next, each of the hidden neuron is connected as a Fully-Connected Neural Network (FNN) with output neurons that form an output vector of indices $\mathbf{o}(i)$, ordered such that the position of vertex index in $\mathbf{o}(i)$ corresponds to the vertex importance. In order to avoid irregularity problem of 3D mesh geometry and topology, we set that each of our hidden neurons is activated by same input weight value and also have same bias for all input neurons. Thus, our algorithm allows for different input sizes and mesh sizes. In [13] we proved that even empirical values of weight coefficients provide good results and thus improve the watermark robustness. The NN learns by adjusting all weights and biases through backpropagation to obtain the ordered vector of vertex indices that provides information of vertex importance. This information will be crucial for host vertex selection in the watermarking task. Now, we describe the learning process of NN and the loss function used in our scheme.

### B. Backpropagation and learning

To illustrate our backpropagation scheme, we consider a hidden neuron $H$ that receives inputs from the input neurons $I$. The activation (output signals) of these neurons are components of the input vector $I = \{\mathbf{i}_i \mid i = 1,\ldots,n\}$ respectively. We denote by $\omega_{IJ}$ the weight between input $I_I$ and our hidden unit $H_J$. For example the net input, $hin_j$ to $j$-th neuron $H$ is the sum of the weighed signals from neurons $\mathbf{i}_1, \mathbf{i}_2, \ldots, \mathbf{i}_n$: $hin_j = b_j + \sum_i \mathbf{i}_i \omega_{ji}$. The bias $b$ is included by adding to the input vector and is treated exactly like any other weight. The activation $h$ of the hidden neuron $H$ is given by some function of its net input, $h = g(hin)$, where the sigmoid function is $\sigma(\mathbf{i}) = (1 + e^{-\mathbf{i}})^{-1}$. The subscripts $JK$ are analogously used for the weight $\omega_{JK}$ between the hidden unit $H_J$ and the output unit $O_K$, which are considered arbitrary, but fixed. With this notation, the corresponding lowercase letters can serve as summation indices in the derivation of the weight update rules. For the arbitrary activation function $g(x)$, its derivative is denoted as $g'$. The dependence of the activation on the weights result from applying the activation function $g$ to the net input $o_K = \sum_j h_j \omega_{jK}$ to find $g(o_K)$. The weights are updated as follows:

#### 1) Updating the hidden layer weights

Let $\mathbf{t} = (t_1, t_2, \ldots, t_n)$ be a training or target output vector. Then, the error to be minimized is $E = (1/2) \sum_k [t_k - o_k]^2$, and using the chain rule [20], we have: $\sigma_K = \partial E / \partial \omega_{JK} = [t_K - o_K] g'(oin_k)$. For weights on connections to the hidden unit $H_J$: $\partial E / \partial \omega_{IJ} = -\sum_k \sigma_k \omega_{Jk} g'(hin_J)[\mathbf{i}_I]$, where $\sigma_J = \sum_k \sigma_k \omega_{Jk} g'(hin_J)$. The gradient weights to the hidden units is $\Delta \omega_{ij} = \eta \sigma_j \mathbf{i}_i$. Since each of our hidden layer neurons are weighted by same weights and biases from all input neurons, the gradient weight to hidden layer is given as $\Delta \omega_j = \eta \sigma_j \mathbf{i}$.

#### 2) Updating the output layer weights

To calculate the output-layer weights, we use standard FNN scheme which minimizes the sum of squares of the errors for all output units: $E = (1/2) \sum_k (h_l - o_k)^2$. Next, we should determine the direction in which to change weights. We calculate the negative error gradient $\nabla E$ with respect to the weights $\omega_{kj}$ and adjust the values of weights. Assume that the activation of this node is equal to the next network input, then output of the $j$-th neuron is $iin_j = g_j(hin_j)$, and thus activation for the output neurons are: $oin_k = \sum_j \omega_{kj} iin_j + b_k$, where our activation output is $o_k = g_k(oin_k)$. As far as the magnitude of weight change is concerned, we take it to be proportional to the negative gradient: $\partial E / \partial \omega_{kj} = (h_k - o_k) g'_k(hin_k) oin_j$. Let $\eta$ is a learning rate used to control the amount of weight adjustment at each step of training. Then, the weights on the output layer are updated according to $\omega_{kj}(t+1) = \omega_{kj}(t) + \Delta\omega_{kj}(t)$, where, $\Delta\omega_{kj} = \eta(h_k - o_k)g'_k(hin_k)iin_j$, and for the sigmoid function weight update is $\omega_{kj}(t+1) = \omega_{kj}(t) + \eta\sigma_k iin_j$, where $\sigma_k = (h_k - o_k)g'_k(oin_k)$.

### C. Training input

According to our input rule we can use any three dimensional training vector regardless of number of its elements. However, the watermark data size determines the number of required host vertices, and thus our neural network may learn from training vectors what size is equal or bigger than the number of watermark bits that should be embedded. On the other hand, our watermark algorithm achieves the watermark capacities that overcomes mentioned problem.

## V. WATERMARKING SCHEME

After the powerful Neuro-OSVETA vertex extraction, as a proven watermarking scheme we use the Sparse variant of Quantization Index Modulation (QIM) technique [21] to embed LDPC coded watermark bits into geometrical structure of 3D mesh. Detailed technique is well explained in our previous work [22]. This section briefly explains the watermarking algorithm used in following three steps.

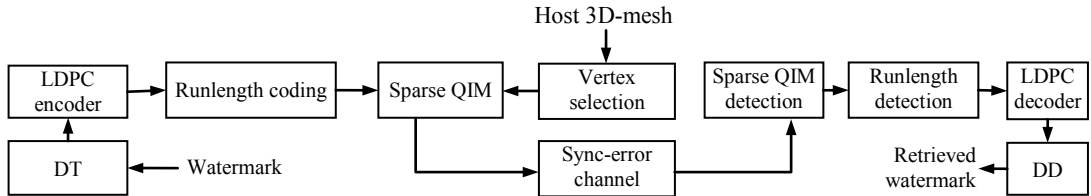

**Figure 3.** A block diagram of the watermarking system

*Step 1).* Firstly, watermarking signal is encoded by structured LDPC codes. Note that there are many good classes of LDPC codes, and our method does not critically depend on which one is used. Here for simplicity, we use regular quasi-cyclic codes, The column weight and row weight of the parity check matrix are $d_v = \mu$ and $d_c = \eta$, respectively. However, before encoding, the watermark sequence is first processed in the *Distribution Transformer* (DT) to optimize probability of zeros and ones, prior to LDPC coding and conversion to runs and transmission through the channel.

*Step 2).* In order to transform the channel with infinite memory into a memoryless channel, the symbols at the input of the channel are represented by runs of bits. The *runlengths* representing zeros and ones are selected according to channel synchronization statistics. In other words, input symbols with odd indices are encoded in runs of $k$ 1's, with $k = 2$ for symbol 0 and $k = 3$ for symbol 1. Symbols with even indices are similarly encoded by runs of 0's.

*Step 3).* Due to a small error in the Euclidean sense may be perceived by the *Human Visual System* (HVS) as a large distortion, the general version of the QIM may lead to serious degradation of the quality of the watermarked 3D signal. To avoid this problem, we use a Sparse variant of QIM. Sparse QIM spreads out the watermark bit over $L$ elements of cover signal **x**. The cover sequence $\mathbf{x}_L$ of length is projected to a $L$-dimensional vector **p** of the unit norm, and the norm of the corresponding projection is quantized. The resulting $L$-dimensional watermarked vector $\mathbf{y}_L$ can be written as $\mathbf{y}_L = \mathbf{x}_L + \left(Q_u(\mathbf{x}_L^T\mathbf{p}) - \mathbf{x}_L^T\mathbf{p}\right)\mathbf{p}$. The detector projects the received watermarked cover vector $\mathbf{r}_L$ to $\mathbf{p}$ and recovers the embedded bit as $\hat{u} = \arg\min_{u \in \{0,1\}} \left\| \mathbf{r}_L^T \mathbf{p}_L - Q_u(\mathbf{r}_L^T \mathbf{p}_L) \right\|_2$. Generally, robustness of the watermark increases with $L$. The principal component axis is the eigenvector that corresponds to the largest eigenvalue of the vertex coordinate covariance matrix.

## VI. Preliminary results

As we already mentioned, there are no existing schemes for NN watermarking, and available training databases with 3D models are used for segmentation and model recognizing purpose. Thus, this paper relies mostly on theoretical assumptions and experimental calculations using dataset used in OSVETA calculations that is equally efficient for small and large 3D models with complex geometrical and topological structures. Our dataset is obtained and derived from 3D models using optimization and simplification functions, as well as using Gaussian filters, which all simulate errors in a 3D mesh structure. Due to our unique requirement, desired resulting output vectors for training set is also constructed using tracing algorithm and semantic correlation to geometric and topologic features in accordance to watermark blindness and robustness.

### A. Hidden layer features

Shape definition of 3D meshes is nontrivial, and it can be shown that many characteristics have influence to the mesh characterization. However, some features can describe the mesh shapes and also perceptual experience of the mesh geometry. For example, selection of important mesh vertices in dependence to Gaussian curvature ($\kappa_G > 0$) and using two different criteria is shown in Figure 4.

We note that the vertices selected by each criterion are at different mesh regions. It can be shown that rest of the features given in Table 1 select different important regions and vertices that can be used for inserting watermark data. Moreover, other criteria listed in [13] can also be used to remove all unimportant vertices and regions from the watermark and thus, decrease the number of input and output neurons which in turn increase the speed and efficiency of our learning process.

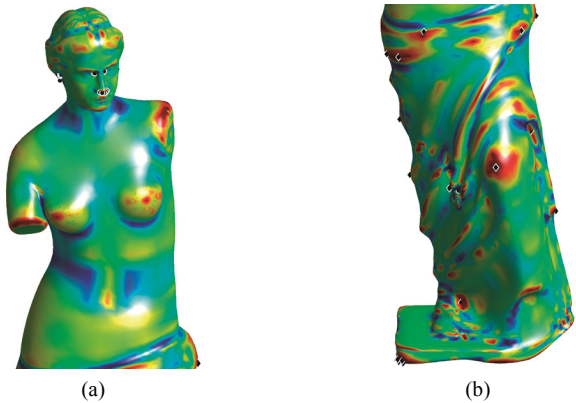

(a)          (b)

**Figure 4.** 3D position of first 50 vertices selected by $\kappa_G$ and $\kappa_{GI}$ criteria: red areas contain vertices with high value of curvature, green areas contain vertices with low value of curvature; (a) $\kappa_G > 0$ - circles, (b), $\kappa_{GI} > 0$ - diamonds.

## B. Probability of vertex deletion

The result of the vertex stability computation in relation to the optimization process is obtained by Neuro-OSVETA and OSVETA. In order to remove topological errors, our model is optimized using 'Pro Optimizer' modifier from 3D Studio Max 2015 application [24] that is based on simplification algorithms. Even with the relatively small training set for our NN, the experimental tests of stability for 1000 vertices selected by Neuro-OSVETA algorithm compared to the OSVETA allocated group of 1000 vertices showed the superiority of our new approach. The results are summarized in Table 2.

**Table 2** The number of vertices deleted by optimization

|  | 0% | 20% | 40% | 60% | 80% | 90% |
|---|---|---|---|---|---|---|
| Total VR | 17350 | 12209 | 6953 | 3926 | 2315 | 1448 |
| Random | 0 | 332 | 622 | 781 | 872 | 920 |
| OSVETA | 0 | 1 | 30 | 147 | 332 | 522 |
| Neuro-OSVETA | 0 | 0 | 22 | 121 | 282 | 421 |

The total number of vertices remaining after simplification (Total RV) with a given percent of remained vertices is given in the first row. Compared to randomly selected vertices (the second row), the third and fourth rows show the number of removed vertices out of 1000 vertices selected by OSVETA and by the Neuro-OSVETA respectively. Probability of vertex deletion as a function of the percent of deleted vertices is shown in Figure 5.

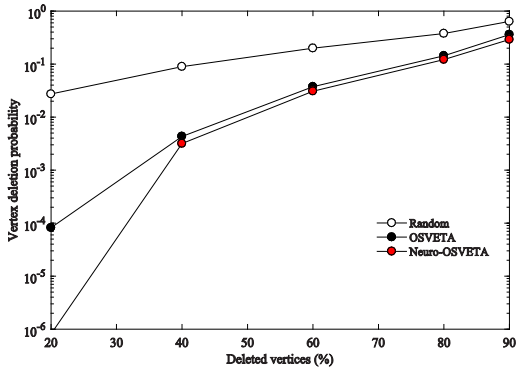

**Figure 5.** Probability of vertex deletion $p_d$ as a function of the deleted vertices.

In Figure 5, our algorithm (shown using red dots) shows better results than the standard OSVETA and much better than randomly selected vertices scheme. Note that this improvement is small due to the small training vector size and can increment using bigger training vector.

## VII. CONCLUSION AND FUTURE WORKS

In this paper, we presented a novel technique that offers a systematic method for vertex selection based on neural networks replacing a heuristic approach in the OSVETA. The Neuro-OSVETA enables a more precise mesh geometry estimation and better curvature and topological feature estimation. From our experimental results, we observe a significant reduction of deletion probability using more accurate identification of stable vertices. Our future works include optimization of NN parameters and using a bigger set of different 3D models as training vector to improve overall performance gain.